\documentclass[conference]{IEEEtran}
\IEEEoverridecommandlockouts

\usepackage{cite}
\usepackage{amsmath,amssymb,amsfonts}
\usepackage{algorithmic}
\usepackage{graphicx}
\usepackage{textcomp}
\usepackage{xcolor}
\usepackage{booktabs}
\usepackage{siunitx}
\usepackage{url}
\usepackage[hidelinks]{hyperref}

\graphicspath{{figures/}}

\newcommand{\methodname}{\textsc{WF-CNN}}

\newcommand{\vect}[1]{\mathbf{#1}}

\def\BibTeX{{\rm B\kern-.05em{\sc i\kern-.025em b}\kern-.08em
    T\kern-.1667em\lower.7ex\hbox{E}\kern-.125emX}}

\begin{document}

\title{Local Order Matters: A Lightweight CPU-Trainable CNN for Website
Fingerprinting on Defended Tor Traffic}

\author{
  \IEEEauthorblockN{Puyang Xu\textsuperscript{*}\thanks{*Corresponding author}}
  \IEEEauthorblockA{
    \textit{Department of Electrical and}\\
    \textit{Computer Engineering}\\
    \textit{Duke University}\\
    Durham, NC 27708, United States\\
    rivocxu@gmail.com
  }
  \and
  \IEEEauthorblockN{Zhizhao Wen}
  \IEEEauthorblockA{
    \textit{Department of Computer Science}\\
    \textit{Rice University}\\
    Houston, TX 77005, United States\\
    zhizhaowen@alumni.rice.edu
  }
  \and
  \IEEEauthorblockN{Yifan Jiang}
  \IEEEauthorblockA{
    \textit{Department of Electrical and}\\
    \textit{Computer Engineering}\\
    \textit{Duke University}\\
    Durham, NC 27708, United States\\
    yifan.jiang999@gmail.com
  }
  \and[\hfill\mbox{}\par\medskip\mbox{}\hfill]
  \IEEEauthorblockN{Chao Wang}
  \IEEEauthorblockA{
    \textit{Department of Computer Science}\\
    \textit{Rice University}\\
    Houston, TX 77005, United States\\
    zylj2020@outlook.com
  }
  \and
  \IEEEauthorblockN{Yusi Gong}
  \IEEEauthorblockA{
    \textit{Department of Computer Science}\\
    \textit{Rice University}\\
    Houston, TX 77005, United States\\
    gyusi73199@gmail.com
  }
}

\maketitle

\begin{abstract}
Website fingerprinting (WF) attacks let a passive adversary infer which site a
Tor user visits from the encrypted cell sequence. On undefended traffic,
hand-crafted-feature attacks are already near-ceiling, so deep learning offers
only a marginal advantage there. We show that the decisive regime is
\emph{defended} traffic: against the FRONT padding defense, which injects dummy
cells that corrupt the volume and count features feature attacks rely on, a
compact \num[group-separator={,}]{95103}-parameter 1-D convolutional network trained on the raw
cell-direction sequence---on CPU only---stays robust while three implemented
direction-only feature baselines degrade. On the closed-world WFlib FRONT set
(95 sites), our model reaches $54.2\%$ top-1 accuracy versus $38.5\%$ for the
strongest implemented feature baseline (k-FP-style RF), a gain of $15.7$
percentage points (\num{3} seeds; seed-0 paired-bootstrap $95\%$ CI
$[+14.5, +17.0]$ pp on per-trace correctness, excluding zero). On
undefended traffic the CNN reaches $94.1\%$ versus $92.3\%$ for k-FP-style RF.
A controlled order-ablation permutes only nonzero traffic cells within each
trace, preserving direction counts, retained length, and trailing zero-padding
positions while destroying traffic order. This collapses the CNN below every feature baseline (from
$54.2\%$ to $7.6\%$, a $46.6$ pp drop), showing that the model depends strongly
on order-derived information under FRONT.
\end{abstract}


\section{Introduction}
\label{sec:introduction}

Tor anonymizes web browsing by routing fixed-size cells through a sequence of
encrypting relays, hiding both the content and the destination of a
connection~\cite{dingledine2004tor}. Website fingerprinting (WF) is the
traffic-analysis attack that undermines this protection: a passive adversary
who observes only the encrypted cell stream between a client and the Tor entry
guard tries to infer which monitored site was visited, using only metadata such
as packet direction, volume, and timing.
WF is a central case study in the systems-security community because it pits a
cheap, deployable observer against a deployed privacy system, and because the
attacker's input is a long, high-dimensional sequence whose discriminative
structure is not obvious a priori.

For roughly a decade the dominant WF attacks combined hand-crafted features with
classical machine learning: statistical and burst features fed to a random
forest (k-FP~\cite{hayes2016kfp}), a cumulative packet-length representation fed
to an RBF-kernel SVM (CUMUL~\cite{panchenko2016cumul}), or thousands of
direction features fed to a weighted k-nearest-neighbor classifier
(Wang-kNN~\cite{wang2014knn}). Deep-learning attacks then learned features
directly from the raw direction sequence~\cite{rimmer2018awf,sirinam2018df}.
On \emph{undefended} closed-world traffic, however, the gap is small: the
strongest feature attacks already reach roughly 95--97\%~\cite{rimmer2018awf,
sirinam2018df}, so the deep-learning advantage there is marginal, and a paper
claiming ``CNN beats features'' on undefended traffic largely re-demonstrates a
known result. We state this honestly rather than inflate it.

The interesting and winnable regime is \emph{defended} traffic. Padding
defences such as FRONT~\cite{gong2020front} inject dummy cells that inflate
and randomize trace volume at zero added latency. This directly corrupts the count- and
volume-based features that k-FP, CUMUL, and Wang-kNN depend on, while a
convolutional model operating on the local arrangement of real and dummy cells
remains comparatively robust. Yet most deep WF results in this regime are
obtained with large, GPU-trained networks~\cite{sirinam2018df,bhat2019varcnn,
rahman2020tiktok}, leaving open whether the defended-traffic advantage survives
in a lightweight, CPU-only attacker and, more importantly, \emph{why} it exists.
Resource requirements are a practical concern in deep-learning workloads,
motivating GPU-memory demand prediction~\cite{wang2025gpumemory}.

We address both questions. We train a compact 1-D CNN (\num[group-separator={,}]{95103} parameters,
roughly $3$--$4\times$ smaller than the Deep Fingerprinting network) on the raw
$\{-1,0,+1\}$ cell-direction sequence, using CPU only, and compare it against
lightweight direction-only proxies inspired by k-FP, CUMUL, and Wang-kNN on the
\emph{same} data and splits from the WFlib benchmark~\cite{deng2024wflib}. These
proxies are not claimed as complete replications of the canonical attacks. To
probe the CNN's dependence on order-derived information, we run an
\emph{order-ablation}: we permute each trace's nonzero traffic prefix, preserving
direction counts and trailing padding positions while destroying traffic order.
The test probes order beyond counts and retained length, without isolating one
unique positional statistic.

\noindent The contributions of this work are:
\begin{itemize}
  \item A lightweight, CPU-trainable 1-D CNN (\num[group-separator={,}]{95103} parameters) for
  closed-world WF on the raw cell-direction sequence, deliberately positioned in
  a different resource regime from GPU-scale WF networks rather than competing
  with them.
  \item A decisive defended-traffic win: on the FRONT-defended WFlib closed
  world, the CNN beats the strongest implemented feature baseline (k-FP-style
  RF) on top-1 accuracy by $15.7$ percentage points across \num{3} seeds, with
  a seed-0 paired-bootstrap 95\% CI of $[+14.5, +17.0]$ pp excluding zero. On
  undefended traffic the gap is much smaller (CNN $94.1\%$ vs k-FP-style RF
  $92.3\%$, a $+1.8$ pp gap).
  \item A controlled order-ablation showing that the CNN depends strongly on
  order-derived information: permuting traffic cells while preserving their
  direction multiset and padding mask collapses the CNN below the feature baselines.
  \item Honest regime positioning: timing is withheld symmetrically from the CNN
  and all baselines, both top-1 accuracy and macro-F1 are reported, and we
  state the scope and implementation differences of the feature baselines.
\end{itemize}

\section{Related Work}
\label{sec:related}

\subsection{Feature-based WF attacks}
The pre-deep-learning WF line engineers features from the packet/cell sequence
and feeds them to classical classifiers. Wang-kNN~\cite{wang2014knn} learns
feature weights for nearest-neighbor classification over thousands of traffic
features. CUMUL~\cite{panchenko2016cumul} trains an RBF-SVM on an interpolated
cumulative packet-length representation. k-FP~\cite{hayes2016kfp} uses a random
forest over roughly 150 statistical, burst, and concentration features to
derive leaf-identifier fingerprints for nearest-neighbor matching.
Our direction-only proxies differ from these original pipelines;
Section~\ref{sec:method} documents the deviations.

\subsection{Deep-learning WF attacks}
Rimmer et al.~\cite{rimmer2018awf} (AWF) evaluated SDAE, CNN, and LSTM models
on raw direction sequences. Deep Fingerprinting (DF)~\cite{sirinam2018df}
uses an eight-layer CNN, reaching $98.3\%$ closed-world accuracy and
undermining WTF-PAD. Var-CNN~\cite{bhat2019varcnn} combines dilated
ResNet models of direction and timing with manually extracted features.
Tik-Tok~\cite{rahman2020tiktok} studies timing leakage and combines raw
timestamps with packet directions. DF reports both CPU and GPU training
costs~\cite{sirinam2018df}; CPU execution alone is therefore not a new WF
capability. Our compact model follows the AWF/DF convolutional lineage.

Dahanayaka et al.~\cite{dahanayaka2022dissecting} analyze CNN filter
activations and identify upload/download transitions near trace fronts as
important WF patterns. Relation-CNN~\cite{zou2024relationcnn} combines
explicit traffic-relation features with NFS-CNN to complement learned
representations. These findings provide related background, not proof of
our model's local-order mechanism.

For defended traffic, Shen et al.~\cite{shen2023robust} aggregate directional
packet counts in time slots into a Traffic Aggregation Matrix and train a CNN,
evaluating defenses including FRONT. This representation uses timing, unlike
our direction-only input; it does not establish our local-order mechanism.

\subsection{WF defences}
FRONT~\cite{gong2020front} randomizes the number and timing of dummy packets
near the trace front without delaying real packets.
RegulaTor~\cite{holland2022regulator} regularizes traffic surges and evaluates
protection against attacks alongside bandwidth and latency overhead, including
comparisons with FRONT. Its low-overhead objective concerns traffic overhead,
not CPU training cost.

More recent practical defenses include BRO~\cite{mcguan2024bro}, a
lightweight, zero-delay padding strategy evaluated against FRONT, and
Palette~\cite{shen2024palette}, which anonymizes traffic clusters in real
time and is implemented as a Tor Pluggable Transport for live traffic.

Xiao et al.~\cite{xiao2024defenses} survey WF defenses and re-evaluate them
under revised attacker assumptions and evaluation settings, highlighting that
reported protection depends on the evaluation protocol.

\subsection{Realistic evaluation}
Cherubin et al.~\cite{cherubin2022online} evaluate online WF using genuine Tor
traffic in a true open world, showing why controlled closed-world performance
alone does not establish practical attack effectiveness.
Deng et al.~\cite{deng2026perspectives} evaluate WF across defenses, traffic
drift, multi-tab browsing, and open-world settings, finding degradation across
conditions. These studies contextualize the limits of our fixed,
closed-world FRONT evaluation.

\section{Method}
\label{sec:method}

\subsection{Problem Formulation}
We study closed-world website fingerprinting as balanced multi-class
classification. A trace is observed as a sequence of Tor cells; in the WFlib
encoding each cell is stored as $X = \text{direction} \times \text{timestamp}$,
from which we recover the \emph{direction sequence}
$\vect{d} = \operatorname{sign}(X) \in \{-1, 0, +1\}^L$, where $+1$ denotes an
outgoing cell, $-1$ an incoming cell, and $0$ padding to the fixed length $L$.
Given $\vect{d}$, the attacker learns a classifier
$f_\theta : \{-1,0,+1\}^L \to \{1,\dots,C\}$ over $C = 95$ monitored sites that
minimizes the average cross-entropy
\begin{equation}
  \mathcal{L}(\theta) = \frac{1}{N} \sum_{i=1}^{N}
  \ell\bigl(f_\theta(\vect{d}_i),\, y_i\bigr),
  \label{eq:loss}
\end{equation}
where $N$ is the number of training traces, $\vect{d}_i$ the $i$-th direction
sequence, $y_i$ its site label, and $\ell$ the cross-entropy loss. We use the
direction sequence only and \emph{withhold timing} ($|X|$) from every method, so
the CNN and all feature baselines see the same information (Section
\ref{sec:experiments}).

\subsection{Compact 1-D CNN}
Our model, \methodname{}, is a compact one-dimensional CNN in the AWF/DF
lineage~\cite{rimmer2018awf,sirinam2018df}, shrunk for CPU training. The input
is the direction sequence reshaped to a single channel of length $L$
($1 \times L$). It applies three convolutional blocks followed by a global
pooling head:
\begin{itemize}
  \item Block 1: \texttt{Conv1d}$(1\!\to\!32, k\!=\!8, \text{stride}\!=\!4)$,
  BatchNorm, ReLU, MaxPool$(8)$;
  \item Block 2: \texttt{Conv1d}$(32\!\to\!64, k\!=\!8)$, BatchNorm, ReLU,
  MaxPool$(4)$;
  \item Block 3: \texttt{Conv1d}$(64\!\to\!128, k\!=\!8)$, BatchNorm, ReLU,
  MaxPool$(4)$;
\end{itemize}
All convolutions use zero padding $4$ per side and dilation $1$, with strides $4/1/1$.
MaxPool strides are $8/4/4$ (equal to their kernels), with padding $0$ and
floor-mode outputs. No layer changes across $L=1000/2000/5000$; at $L=1000$,
successive conv/pool lengths are $251/31/32/8/9/2$.
The head uses adaptive average pooling to length $1$, dropout with rate $0.3$, and a
linear layer to the $C = 95$ class logits. The stride-4 first convolution and
the global-average-pooling head (rather than DF's two $512$-unit fully connected
layers) keep the parameter count to \num[group-separator={,}]{95103}, roughly $3$--$4\times$ smaller
than the DF network, and the model trains on CPU. The small convolutional
kernels operate over the \emph{local} arrangement of real and dummy cells, which
is central to the mechanism we test below.

\subsection{Feature Baselines}
We compare against three scalable, direction-only feature baselines on the
\emph{same} data and splits. They are inspired by canonical WF attacks but are
implementation-specific proxies rather than complete replications:
\begin{itemize}
  \item \textbf{k-FP-style RF}~\cite{hayes2016kfp}: a 53-dimensional
  direction-derived subset containing global counts and fractions, burst
  summaries, outgoing-cell concentration over 20 chunks, and cumulative
  outgoing counts at fixed trace fractions. Standardized features are fed to a
  200-tree random forest.
  \item \textbf{CUMUL-RF}~\cite{panchenko2016cumul}: four volume/count values
  plus 100 interpolated cumulative-sum points (104 dimensions), standardized
  and classified by a 200-tree random forest. This retains the CUMUL
  representation but replaces the original RBF-SVM for full-data CPU
  scalability.
  \item \textbf{Wang-kNN-style}~\cite{wang2014knn}: the same standardized
  53-dimensional direction-derived vector used by k-FP-style RF, classified by
  distance-weighted k-NN with $k = 5$. It is not the original Wang feature and
  distance pipeline.
\end{itemize}
A random/majority floor ($1/95 \approx 1.05\%$) is reported for sanity. These
baselines contain both count summaries and some order-derived burst,
concentration, or cumulative features; our comparisons therefore apply only to
the implementations above.

\subsection{Order-Ablation Design}
We retrain after shuffling only each trace's nonzero prefix, including FRONT
dummy cells ($\pm1$). Auditing all FRONT arrays confirmed positive, nondecreasing
timestamp magnitudes followed only by zeros; asserted prefix/suffix masks
therefore recover retained lengths. Direction counts, lengths, and padding
positions stay fixed, correcting the earlier full-vector shuffle's padding
redistribution. Train/validation/test permutations are drawn once with seeds
$s/s{+}1/s{+}2$; splits, architecture, and training match full-order seed $s$.
Multiple order-dependent structures still change together: this neither isolates
a unique local-order mechanism nor establishes feature-baseline invariance.

\section{Experiments}
\label{sec:experiments}

\subsection{Setup}
\textbf{Dataset and sampling.} WFlib~\cite{deng2024wflib}
(Zenodo DOI 10.5281/zenodo.13732130, CC-BY-4.0) provides 95,000 FRONT traces
(1,000/site) and 105,730 CW traces (1,051--1,125/site), with 95 class labels.
We use a fixed 500-trace-per-site protocol selected during preliminary CPU-only
convergence checks and apply it independently to each dataset. Per seed
$s\in\{0,1,2\}$, one NumPy \texttt{default\_rng(s)} shuffles indices within
classes in ascending class order; the first 500 become 350/75/75
train/validation/test instances ($70/15/15$). Of 47,500 selected traces/seed,
all methods share the splits; the remaining 47,500 FRONT or 58,230 CW traces
are unused. Subsets are redrawn across seeds and may overlap. Thus, CW/FRONT
share sample counts, seeds, and protocol, but no trace-level correspondence is
assumed. Initialization and baseline randomness also use $s$.
We take $\operatorname{sign}(X)$ of the stored direction-times-timestamp arrays
and retain the first $L=5000$ entries, including existing trailing zero padding.
Timing ($|X|$) is withheld from every method; timing-aware
attacks~\cite{rahman2020tiktok} remain out of scope.

\textbf{Metrics and significance.} The primary metric is closed-world top-1
accuracy on the balanced 95-class test set; we also report macro-F1. The
headline significance statistic is a paired bootstrap 95\% confidence
interval on the aligned per-trace correctness difference between the CNN and
the strongest feature baseline for seed 0 (10,000 resamples); an interval
excluding zero indicates a significant improvement for that run. Separately,
we report the 3-seed mean $\pm$ sample standard deviation and the mean of the
three seed-level paired margins as consistency checks.

\textbf{Compute.} All models are trained and evaluated on CPU only (\num{4}
threads). The CNN uses Adam (learning rate $10^{-3}$, batch size 256),
cross-entropy loss, and at most 40 epochs, restoring the best-validation model.
ReduceLROnPlateau uses factor $0.5$, patience 4, and minimum rate $10^{-5}$;
early stopping uses patience 12. The corrected ablation uses the same settings.

\subsection{Baselines}
We compare \methodname{} against the k-FP-style RF, CUMUL-RF, and
Wang-kNN-style proxies described in Section~\ref{sec:method}, all run on
identical direction-only data and splits. In addition to withholding timing,
the implementations deliberately use compact feature sets and scalable
classifiers: 53 rather than the full k-FP feature pipeline, RandomForest rather
than the original CUMUL RBF-SVM, and the shared 53-dimensional vector rather
than the original Wang feature/distance pipeline. A random/majority floor
($\approx 1.05\%$) is included for sanity.

\subsection{Main Results}
Table~\ref{tab:main} reports closed-world top-1 accuracy and macro-F1 (mean
$\pm$ std over \num{3} seeds) for each attack on the FRONT-defended set. On
FRONT, \methodname{} reaches $54.2\% \pm 0.4\%$ top-1 accuracy versus
$38.5\% \pm 0.4\%$ for the strongest implemented feature baseline (k-FP-style
RF), a gain of $+15.7$ percentage points. The seed-0 paired-bootstrap 95\% CI
on per-trace correctness is $[+14.5, +17.0]$ pp, excluding zero
($P(\text{CNN}\!>\!\text{k-FP}) = 1.0$); the mean of the three seed-level
paired margins is $+15.7$ pp. \methodname{} also beats Wang-kNN-style by
$+23.5$ pp and CUMUL-RF by $+29.2$ pp. Macro-F1 tracks accuracy (CNN $53.7\%$
vs k-FP-style RF $37.1\%$, $+16.6$ pp), so the
accuracy/F1 divergence sometimes reported under FRONT is small in this
benchmark and does not undercut the headline.

\begin{table}[!tb]
\caption{FRONT-defended closed-world attack accuracy and macro-F1 (95 sites, mean $\pm$ std, 3 seeds; best in bold).}
\label{tab:main}
\centering
\setlength{\tabcolsep}{4pt}
\begin{tabular}{lcc}
\toprule
Method & Accuracy & Macro-F1 \\
\midrule
Wang-kNN & $0.307 \pm {<}0.001$ & $0.306 \pm 0.001$ \\
CUMUL & $0.250 \pm 0.005$ & $0.239 \pm 0.005$ \\
k-FP (RF) & $0.385 \pm 0.004$ & $0.371 \pm 0.003$ \\
CNN (ours) & $\boldsymbol{0.542 \pm 0.004}$ & $\boldsymbol{0.537 \pm 0.004}$ \\
\bottomrule
\end{tabular}
\end{table}

\begin{figure}[!tb]
  \centering
  \includegraphics[width=\columnwidth]{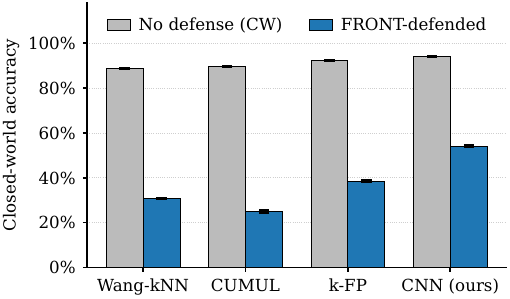}
  \caption{Closed-world accuracy under No-defense vs.\ FRONT (mean of 3
  seeds, $\pm 1$ std). The implemented direction-only feature baselines degrade under FRONT;
  \methodname{} stays robust.}
  \label{fig:bars}
\end{figure}

On the undefended set (CW, same class counts and sampling/split protocol), every
method performs strongly: CNN $94.1\% \pm 0.1\%$, k-FP-style RF
$92.3\% \pm 0.3\%$, Wang-kNN-style $88.8\% \pm 0.3\%$, and CUMUL-RF
$89.6\% \pm 0.2\%$. The CNN's edge over k-FP-style RF shrinks from $+15.7$ pp
under FRONT to only $+1.8$ pp on CW. This is a cross-dataset contrast, not a
paired estimate of the defense's causal effect (Fig.~\ref{fig:bars}).
From CW to FRONT, k-FP-style RF falls from $92.3\%$ to
$38.5\%$ ($-53.8$ pp), Wang-kNN-style from $88.8\%$ to $30.7\%$ ($-58.1$ pp),
and CUMUL-RF from $89.6\%$ to $25.0\%$ ($-64.6$ pp), while \methodname{} falls
from $94.1\%$ to $54.2\%$ ($-39.9$ pp).

Across the three seeds the CNN's test accuracy has sample standard deviation
$0.4$ pp, and validation accuracy generally improves over the 40 CPU epochs
under the ReduceLROnPlateau schedule.

\subsection{Ablation Study}
Table~\ref{tab:ablation} reports the order-ablation and an input-length sweep.

\emph{Order-ablation.} Full-order \methodname{} reaches $54.2\%$ on FRONT, but
permuting only the nonzero traffic prefix while fixing all padding positions
reduces accuracy to $7.6\%\pm0.1\%$ (three-seed mean and sample standard
deviation)---a $46.6$ pp drop. This places the shuffled CNN \emph{below} every implemented
feature baseline (k-FP-style RF $38.5\%$, Wang-kNN-style $30.7\%$, CUMUL-RF
$25.0\%$). The result shows that preserved direction counts and padding layout are
insufficient for the CNN and that order-derived information is important; it
does not identify local order as the unique causal mechanism.

\emph{Input-length sweep.} Truncating the input from $5000$ to $2000$ cells
drops accuracy from $54.6\%$ to $41.7\%$; at $L\!=\!1000$ it falls to $27.6\%$
and the CNN no longer beats k-FP. The defended-traffic win therefore depends
on the model seeing enough of the trace to recover the surviving local
ordering, consistent with the mechanism above. The headline result at
$L\!=\!5000$ is single-seed in this sweep; the $L\!=\!5000$ row in the main
table is the $3$-seed mean and lies within $0.5$ pp.

\begin{table}[!tb]
\caption{FRONT-defended ablations: order-ablation (top) and input-length sweep (bottom). Top rows are 3-seed means; length-sweep rows are single-seed.}
\label{tab:ablation}
\centering
\setlength{\tabcolsep}{4pt}
\begin{tabular}{lc}
\toprule
Configuration & Accuracy \\
\midrule
CNN (ours), full order & $0.542 \pm 0.004$ \\
CNN, shuffled (padding fixed) & $0.076 \pm 0.001$ \\
best feature baseline (k-FP) & $0.385 \pm 0.004$ \\
\midrule
CNN, input length 1000 & $0.276$ \\
CNN, input length 2000 & $0.417$ \\
CNN, input length 5000 & $0.546$ \\
\bottomrule
\end{tabular}
\end{table}

\subsection{Discussion}
\label{sec:discussion}
The win is on the pre-committed primary metric (top-1 accuracy); the $+15.7$ pp
margin is large relative to seed dispersion and the seed-0 paired-bootstrap CI
excludes zero. The corrected order-ablation preserves direction counts, retained
length, and padding positions while destroying traffic order, reducing the CNN below the
feature baselines. Together with the CNN-vs.-k-FP-style gap shrinking from
$+15.7$ pp under FRONT to $+1.8$ pp on the undefended set, this supports
an interpretation that order-derived information matters especially under
FRONT; it does not prove a unique mechanism or a generic advantage over every
canonical feature attack.

\textbf{Scope and limitations.} We target the lightweight CPU regime
(\num[group-separator={,}]{95103} parameters, $3$--$4\times$ smaller than DF~\cite{sirinam2018df})
and do not compete with GPU/timing-augmented WF networks such as
Var-CNN~\cite{bhat2019varcnn} or Tik-Tok~\cite{rahman2020tiktok}; timing is
withheld symmetrically from every method. The evaluation is closed-world on a
single defence (FRONT). The three feature baselines are scalable direction-only
proxies: k-FP-style RF uses 53 features rather than the full canonical pipeline,
CUMUL-RF replaces the original RBF-SVM, and Wang-kNN-style reuses the same
53-dimensional vector rather than the original Wang features and distance.
Consequently, the results compare against these implementations and should not
be read as definitive comparisons with the complete canonical attacks.

\section{Conclusion}
\label{sec:conclusion}
A \num[group-separator={,}]{95103}-parameter, CPU-trainable 1-D CNN on the raw Tor cell-direction
sequence reaches $54.2\%$ closed-world top-1 accuracy on FRONT-defended traffic,
beating the strongest implemented feature baseline (k-FP-style RF) by $+15.7$
percentage points, with the seed-0 paired-bootstrap 95\% CI on per-trace
correctness fully excluding zero. The gap is only $+1.8$ pp on
undefended traffic, so the advantage is specific to this defended setting. A
controlled within-trace permutation that preserves direction counts and padding positions
collapses the CNN to $7.6\%$, below every implemented feature baseline,
demonstrating strong dependence on order-derived information without claiming a
unique causal mechanism.
The evaluation is closed-world on FRONT alone; multi-defence (WTF-PAD,
Walkie-Talkie) and open-world studies remain future work.

\bibliographystyle{IEEEtran}
\bibliography{references}

@inproceedings{hayes2016kfp,
  author    = {Jamie Hayes and George Danezis},
  title     = {{k-fingerprinting}: A Robust Scalable Website Fingerprinting Technique},
  booktitle = {25th {USENIX} Security Symposium ({USENIX} Security 16)},
  pages     = {1187--1203},
  year      = {2016},
  publisher = {{USENIX} Association},
  url       = {https://www.usenix.org/conference/usenixsecurity16/technical-sessions/presentation/hayes},
}

@inproceedings{panchenko2016cumul,
  author    = {Andriy Panchenko and Fabian Lanze and Andreas Zinnen and Martin Henze and Jan Pennekamp and Klaus Wehrle and Thomas Engel},
  title     = {Website Fingerprinting at Internet Scale},
  booktitle = {23rd Annual Network and Distributed System Security Symposium ({NDSS} 2016)},
  year      = {2016},
  doi       = {10.14722/ndss.2016.23477},
  url       = {https://doi.org/10.14722/ndss.2016.23477},
}

@inproceedings{wang2014knn,
  author    = {Tao Wang and Xiang Cai and Rishab Nithyanand and Rob Johnson and Ian Goldberg},
  title     = {Effective Attacks and Provable Defenses for Website Fingerprinting},
  booktitle = {23rd {USENIX} Security Symposium ({USENIX} Security 14)},
  pages     = {143--157},
  year      = {2014},
  publisher = {{USENIX} Association},
  url       = {https://www.usenix.org/conference/usenixsecurity14/technical-sessions/presentation/wang_tao},
}

@inproceedings{rimmer2018awf,
  author    = {Vera Rimmer and Davy Preuveneers and Marc Juarez and Tom Van Goethem and Wouter Joosen},
  title     = {Automated Website Fingerprinting through Deep Learning},
  booktitle = {25th Annual Network and Distributed System Security Symposium ({NDSS} 2018)},
  year      = {2018},
  doi       = {10.14722/ndss.2018.23105},
  url       = {https://doi.org/10.14722/ndss.2018.23105},
}

@inproceedings{sirinam2018df,
  author    = {Payap Sirinam and Mohsen Imani and Marc Juarez and Matthew Wright},
  title     = {Deep Fingerprinting: Undermining Website Fingerprinting Defenses with Deep Learning},
  booktitle = {Proceedings of the 2018 {ACM} {SIGSAC} Conference on Computer and Communications Security ({CCS} '18)},
  pages     = {1928--1943},
  year      = {2018},
  publisher = {{ACM}},
  doi       = {10.1145/3243734.3243768},
  url       = {https://doi.org/10.1145/3243734.3243768},
}

@article{bhat2019varcnn,
  author    = {Sanjit Bhat and David Lu and Albert Kwon and Srinivas Devadas},
  title     = {{Var-CNN}: A Data-Efficient Website Fingerprinting Attack Based on Deep Learning},
  journal   = {Proceedings on Privacy Enhancing Technologies ({PoPETs})},
  volume    = {2019},
  number    = {4},
  pages     = {292--310},
  year      = {2019},
  doi       = {10.2478/popets-2019-0070},
  url       = {https://doi.org/10.2478/popets-2019-0070},
}

@article{rahman2020tiktok,
  author    = {Mohammad Saidur Rahman and Payap Sirinam and Nate Mathews and Kantha Girish Gangadhara and Matthew Wright},
  title     = {{Tik-Tok}: The Utility of Packet Timing in Website Fingerprinting Attacks},
  journal   = {Proceedings on Privacy Enhancing Technologies ({PoPETs})},
  volume    = {2020},
  number    = {3},
  pages     = {5--24},
  year      = {2020},
  doi       = {10.2478/popets-2020-0043},
  url       = {https://doi.org/10.2478/popets-2020-0043},
}

@inproceedings{dingledine2004tor,
  author    = {Roger Dingledine and Nick Mathewson and Paul Syverson},
  title     = {{Tor}: The Second-Generation Onion Router},
  booktitle = {13th {USENIX} Security Symposium ({USENIX} Security 04)},
  pages     = {303--320},
  year      = {2004},
  publisher = {{USENIX} Association},
  url       = {https://www.usenix.org/conference/13th-usenix-security-symposium/tor-second-generation-onion-router},
}

@inproceedings{gong2020front,
  author    = {Jiajun Gong and Tao Wang},
  title     = {Zero-delay Lightweight Defenses against Website Fingerprinting},
  booktitle = {29th {USENIX} Security Symposium ({USENIX} Security 20)},
  pages     = {717--734},
  year      = {2020},
  publisher = {{USENIX} Association},
  url       = {https://www.usenix.org/conference/usenixsecurity20/presentation/gong},
}

@inproceedings{deng2024wflib,
  author    = {Xinhao Deng and Qi Li and Ke Xu},
  title     = {Robust and Reliable Early-Stage Website Fingerprinting Attacks via Spatial-Temporal Distribution Analysis},
  booktitle = {Proceedings of the 2024 {ACM} {SIGSAC} Conference on Computer and Communications Security ({CCS} '24)},
  year      = {2024},
  publisher = {{ACM}},
  doi       = {10.1145/3658644.3670272},
  url       = {https://doi.org/10.1145/3658644.3670272},
}

@inproceedings{wang2025gpumemory,
  author    = {Chao Wang and Zhizhao Wen and Ruoxin Zhang and Puyang Xu and Yifan Jiang},
  title     = {{GPU} Memory Requirement Prediction for Deep Learning Task Based on Bidirectional Gated Recurrent Unit Optimization Transformer},
  booktitle = {2025 5th International Conference on Artificial Intelligence, Virtual Reality and Visualization ({AIVRV})},
  year      = {2025},
  pages     = {31--35},
  doi       = {10.1109/AIVRV67401.2025.11350369},
  url       = {https://doi.org/10.1109/AIVRV67401.2025.11350369},
}

@inproceedings{shen2023robust,
  author    = {Meng Shen and Kexin Ji and Zhenbo Gao and Qi Li and Liehuang Zhu and Ke Xu},
  title     = {Subverting Website Fingerprinting Defenses with Robust Traffic Representation},
  booktitle = {32nd {USENIX} Security Symposium ({USENIX} Security 23)},
  pages     = {607--624},
  year      = {2023},
  publisher = {{USENIX} Association},
  url       = {https://www.usenix.org/conference/usenixsecurity23/presentation/shen-meng},
}

@article{holland2022regulator,
  author    = {James K. Holland and Nicholas Hopper},
  title     = {{RegulaTor}: A Straightforward Website Fingerprinting Defense},
  journal   = {Proceedings on Privacy Enhancing Technologies ({PoPETs})},
  volume    = {2022},
  number    = {2},
  pages     = {344--362},
  year      = {2022},
  doi       = {10.2478/popets-2022-0049},
  url       = {https://doi.org/10.2478/popets-2022-0049},
}

@article{xiao2024defenses,
  author    = {Xi Xiao and Xiang Zhou and Zhenyu Yang and Le Yu and Bin Zhang and Qixu Liu and Xiapu Luo},
  title     = {A Comprehensive Analysis of Website Fingerprinting Defenses on {Tor}},
  journal   = {Computers \& Security},
  volume    = {136},
  pages     = {103577},
  year      = {2024},
  doi       = {10.1016/j.cose.2023.103577},
  url       = {https://doi.org/10.1016/j.cose.2023.103577},
}

@inproceedings{cherubin2022online,
  author    = {Giovanni Cherubin and Rob Jansen and Carmela Troncoso},
  title     = {Online Website Fingerprinting: Evaluating Website Fingerprinting Attacks on {Tor} in the Real World},
  booktitle = {31st {USENIX} Security Symposium ({USENIX} Security 22)},
  pages     = {753--770},
  year      = {2022},
  publisher = {{USENIX} Association},
  url       = {https://www.usenix.org/conference/usenixsecurity22/presentation/cherubin},
}

@article{deng2026perspectives,
  author    = {Xinhao Deng and Jingyou Chen and Linxiao Yu and Yixiang Zhang and Zhongyi Gu and Changhao Qiu and Xiyuan Zhao and Ke Xu and Qi Li},
  title     = {Beyond a Single Perspective: Towards a Realistic Evaluation of Website Fingerprinting Attacks},
  journal   = {Tsinghua Science and Technology},
  volume    = {31},
  number    = {3},
  pages     = {1381--1392},
  year      = {2026},
  doi       = {10.26599/TST.2025.9010167},
  url       = {https://doi.org/10.26599/TST.2025.9010167},
}

@article{dahanayaka2022dissecting,
  author    = {Thilini Dahanayaka and Guillaume Jourjon and Suranga Seneviratne},
  title     = {Dissecting Traffic Fingerprinting {CNNs} with Filter Activations},
  journal   = {Computer Networks},
  volume    = {206},
  pages     = {108770},
  year      = {2022},
  doi       = {10.1016/j.comnet.2022.108770},
  url       = {https://doi.org/10.1016/j.comnet.2022.108770},
}

@article{zou2024relationcnn,
  author    = {Hongcheng Zou and Ziling Wei and Jinshu Su and Shuhui Chen and Zheng Qin},
  title     = {{Relation-CNN}: Enhancing Website Fingerprinting Attack with Relation Features and {NFS-CNN}},
  journal   = {Expert Systems with Applications},
  volume    = {247},
  pages     = {123236},
  year      = {2024},
  doi       = {10.1016/j.eswa.2024.123236},
  url       = {https://doi.org/10.1016/j.eswa.2024.123236},
}

@article{mcguan2024bro,
  author    = {Colman McGuan and Chansu Yu and Kyoungwon Suh},
  title     = {Practical and Lightweight Defense against Website Fingerprinting},
  journal   = {Computer Communications},
  volume    = {228},
  pages     = {107976},
  year      = {2024},
  doi       = {10.1016/j.comcom.2024.107976},
  url       = {https://doi.org/10.1016/j.comcom.2024.107976},
}

@inproceedings{shen2024palette,
  author    = {Meng Shen and Kexin Ji and Jinhe Wu and Qi Li and Xiangdong Kong and Ke Xu and Liehuang Zhu},
  title     = {Real-Time Website Fingerprinting Defense via Traffic Cluster Anonymization},
  booktitle = {2024 {IEEE} Symposium on Security and Privacy ({SP})},
  year      = {2024},
  publisher = {{IEEE}},
  doi       = {10.1109/SP54263.2024.00247},
  url       = {https://doi.org/10.1109/SP54263.2024.00247},
}

\end{document}